# A Thermal Harmonic Field Description of Phase Transition: The Alternative Approach to the Landau Theory

Yi Wang,[*] Long-Qing Chen, and Zi-Kui Liu

*Materials Science and Engineering, The Pennsylvania State University, University Park, PA 16802, USA.*

The study of critical phenomena and phase transitions is an important part of modern condensed matter physics. In this regard, the phenomenological Landau theory [1] has been extraordinarily useful. Hereby we present an alternative theoretical description to the Landau theory for a system under phase transition, based on *a priori* assumption that the macroscopic system is made of the thermal mixing among multi harmonics each of them can be distinguished by crystal orientation, polar direction, magnetic direction, or even momentum etc. Our theory naturally gives rise to a long range field and is able to account for both the type of lattice and the spatial dimensionality, in addition to that the excess free energy is referenced to the low temperature structure together with the positive excess entropy. The improvements over the Landau theory are demonstrated using ferroelectric-paraelectric system of $PbTiO_3$ on its phase transition and associated thermodynamic behaviors.

[*] Email: yuw3@psu.edu

I. **Introduction**

In developing the theory for describing the critical phenomena and phase transitions, Landau was motivated to suggest that the free energy of any system should obey two conditions: that the free energy is analytic, and that it obeys the symmetry of the Hamiltonian. As a result, the free-energies of all phase transitions in the vicinity of the phase transition are expanded as an even power series as a Taylor expansion in the order parameter. Take the ferroelectric material Lead titanate (PbTiO$_3$) as an example, the excess free-energy, $\Delta G^{LD}$, as a function of spontaneous polarization, $P_s$, is expressed by the Landau theory as [2-3]

$$\text{Eq. 1,  } \Delta G^{LD}(P_s, T) = A_0(T - T_0)P_s^2 + BP_s^4 + CP_s^6$$

where the parameter values are widely taken as $T_0$ = 751.95 K, $A_0$ = 14.315 J/mole K$^{-1}$, $B$ = -2757.4 J/mole, and $C$ = 9908.7 J/mole for PbTiO$_3$ by Haun et al. [3]. Employing these parameter values, the evolutions against $T$ of the isothermal excess free energy as a function of the spontaneous polarization for PbTiO$_3$ are plotted in Figure 1a. PbTiO$_3$ is characterized with its ferroelectric-paraelectric transition [2-4] occurred at ~764 K. Indeed, the critical phenomena of the ferroelectric-paraelectric transition of PbTiO$_3$ are modeled very well, including that for $T$ < 764 K the point at $P_s$ = 0 corresponds to a local maximum of the free energy, and the equilibrium state is one of the two states of spontaneously broken symmetry for which $\Delta G^{LD}$ has an absolute minimum; while for $T$ > 764 K the point at $P_s$ = 0 corresponds to a global minimum of the free energy along the isothermal curves.

However, we may point out that the insufficiency of the Landau theory has been mostly overlooked. It has been also pointed out that "*Unfortunately, the critical exponent values or the*

*quantitative description actually predicted from the Landau expansion are, in most cases, not correct!*". This is demonstrated in Figure 2, by examining the behaviors of the heat capacity, excess entropy, and excess enthalpy when approaching to the phase transition, that illustrates the inaccuracy of the Landau theory in the description of the measurable thermodynamic properties [4-5]. For the convenience of direct comparison with experimental data, in Figure 1 the background phonon contribution from the ferroelectric structure to the thermodynamic quantities has been calculated using the first-principles approach which will be detailed later in this paper. It is clearly seen that, in the intermediate temperature range up to the critical temperature, the Landau model shows substantial discrepancies from the measured heat capacity data [4-5] or the estimated excess entropy or enthalpy data by Rossetti et al. [5] based on their measured heat capacity data.

We might also add that the Landau model is not convenient for a complete thermodynamic simulation. Note the fact that in Landau theory the excess free-energy is referenced to the high temperature high symmetry phase whose free energy, to our limited knowledge, has never been well formulated or calculated within the modern first-principles theory. Take the ferroelectric material of $PbTiO_3$ as an example, the accurate phonon description of the free energy for the high temperature paraelectric phase has been impossible due to the existence of the imaginary phonon modes.

The present work is inspired by *a priori* observation [6] on the common characteristics showed for the ferroelectric, the magnetic (ferromagnetic as well as antiferromagnetic), and the superconductive transitions – 1) the similarity of the temperature evolutions among the polarization of the ferroelectric materials, the magnetization of the magnetic materials, and the pseudo energy gap of the superconductive materials; and 2) the common temperature evolution behaviors of the heat capacity of these phenomena. Since our free energy model detailed below is based on the viewpoint of thermal mixing among multiple harmonics differed by orientations, we abbreviate our approach as the thermal harmonic field (THF) model.

II. **The THF Model**

The theoretical foundation of the THF model is built up on our assumption on the phase transition phenomena in a viewpoint of *dynamic energy fluctuation*. Let us start by considering a macroscopic system made of massive number of small single crystal domains (SSCD) of the same size and each of them is quantized with respect to an orientation order parameter represented by a vector $\mathbf{X}$. We will investigate the energetics of the individual SSCD. We first assume that the individual SSCD possesses the lowest energy at $\mathbf{X} = \mathbf{X}_0(g)$ with the index $g$ labeling the preferred orientation of the considered SSCD. Then, we look at the energy change of the SSCD when its orientation is deviated from its preferred orientation $\mathbf{X}_0(g)$. We write down the energy fluctuation as

Eq. 2. $\Delta F(\mathbf{X}; g) = \frac{1}{2}[\mathbf{X} - \mathbf{X}_0(g)] \cdot \mathbf{\Phi}(g) \cdot [\mathbf{X} - \mathbf{X}_0(g)]$

where $\mathbf{\Phi}(g)$ is a tensor representing the second order derivative of the energy of the SSCD with respect to the vector $\mathbf{X}$, noting that the dot symbol "$\cdot$" in Eq. 2 represents the algebraic operation of dot product defined in linear algebra. Eq. 1 can, in fact, be understood more easily in the sense of mathematical expand of the Hamiltonian of the SSCD with accuracy up to the second order in the Taylor series as one knows the fact that the first order term of $\Delta F(\mathbf{X}; g)$ with respect to $\mathbf{X}$ is zero at $\mathbf{X}_0(g)$ which is an equilibrium value of $\mathbf{X}$. At finite temperature, we advocate that the macroscopic system is a mixture of SSCDs with different orientation $g$. As a result, our theoretical foundation is accomplished by proposing to calculate the excess free-energy of the system with:

$$\text{Eq. 3 } \Delta G(\mathbf{X},T) = -\frac{1}{\beta}\ln\left\{\sum_g \exp[-\beta\Delta F(\mathbf{X};g)]\right\}$$

where $\beta = 1/k_B T$ with $k_B$ being the Boltzmann's constant and $T$ the temperature. Hereby we want to point out that the variable $\mathbf{X}$ is thermodynamically an independent variable from the temperature $T$. The derivations of two important thermodynamic quantities are exemplified below.

The equilibrium value $\mathbf{X}_{eq}$ of $\mathbf{X}$ can be calculated under the condition of $\frac{\partial \Delta G(\mathbf{X},T)}{\partial \mathbf{X}} = 0$ and we get

$$\text{Eq. 4. } \mathbf{X} \cdot \sum_g \mathbf{\Phi}(g)\exp[-\beta\Delta F(\mathbf{X};g)] = \sum_g \mathbf{X}_0(g) \cdot \mathbf{\Phi}(g)\exp[-\beta\Delta F(\mathbf{X};g)]$$

The Curie–Weiss temperature $T_0$ for a phase transition can be calculated under the condition the condition of $\frac{\partial^2 \Delta G(\mathbf{X},T)}{\partial \mathbf{X}\partial \mathbf{X}} = 0$ at $\mathbf{X} = 0$, which gives the critical equation in the tensor form

$$\text{Eq. 5, } k_B T_0 \sum_g \mathbf{\Phi}(g) = \sum_g [\mathbf{\Phi}(g)\cdot\mathbf{X}_0(g)] \otimes [\mathbf{X}_0(g)\cdot\mathbf{\Phi}(g)]$$

where the symbol "$\otimes$" in Eq. 5 represents the algebraic operation of "direct product" defined in linear algebra.

### III.  **The Derived Physics from the THF model**

Now we make in-depth analysis of the physical significance of Eq. 3 and we will see how a good description of the excess free energy can be reached by using at most the second order term of the orientation order parameter, compared with those use up to the sixth order term by the Landau theory. For simplicity, we first write $\mathbf{X}_0(g)$ in the form $\mathbf{X}_0(g) = X_0 \hat{\mathbf{x}}_g$ where $X_0$ is the scalar value of $\mathbf{X}_0(g)$ and $\hat{\mathbf{x}}_g$ is an unit vector representing the direction of $\mathbf{X}_0(g)$; Next we write $\mathbf{X} = xX_0\hat{\mathbf{x}}$ where $x$ is a reduced value representing the relative scalar value of $\mathbf{X}$ to $X_0$ and $\hat{\mathbf{x}}$ is an unit vector representing the direction of $\mathbf{X}$. Together with imposing invariance equality of $\hat{\mathbf{x}}_g \cdot \mathbf{\Phi}(g) \cdot \hat{\mathbf{x}}_g = \hat{\mathbf{x}}_0 \cdot \mathbf{\Phi}(0) \cdot \hat{\mathbf{x}}_0$, Eq. 3 is now simplified to

Eq. 6, $\Delta G(\mathbf{X},T) = \dfrac{k_B T_k}{2}(1+x^2) - \dfrac{1}{\beta}\ln\sum_g \exp\left\{[-\dfrac{x^2}{2}D(g) + x\hat{\mathbf{x}}\cdot\hat{\mathbf{x}}_g]\dfrac{T_k}{T}\right\}$

where we have introduced an effective temperature of $T_k = \hat{\mathbf{x}}\cdot\mathbf{\Phi}(0)\cdot\hat{\mathbf{x}} X_0^2/k_B$ within which the crystal is assumed to be initially orientated at the direction labeled by $g = 0$. The quantity $D(g)$ in Eq. 6 takes the form

Eq. 7, $D(g) = \dfrac{\hat{\mathbf{x}}\cdot\mathbf{\Phi}(g)\cdot\hat{\mathbf{x}}}{\hat{\mathbf{x}}\cdot\mathbf{\Phi}(0)\cdot\hat{\mathbf{x}}} - 1$.

which accounts for the anisotropy of the SSCD and plays an important role of dictating the critical behavior in the proximity of phase transition. It is amazing to note that Eq. 6 naturally brings out an effective *long range field*, $x\hat{\mathbf{x}}T_k/T$, which at the finite temperature modulates the thermal distributions to the excess free energy of all constituent SSCDs as if they are dipoles along the direction of $\hat{\mathbf{x}}_g$'s.

The evaluation of the equilibrium value $x_{eq}$ can be simplified by projecting Eq. 4 into the $\hat{\mathbf{x}}$ direction, resulting in the problem to find the roots of the following one-dimensional equation:

$$\text{Eq. 8 } x = \frac{\sum_g \hat{\mathbf{x}} \cdot \hat{\mathbf{x}}_g \exp\left\{[-\frac{x^2}{2}D(g) + x\hat{\mathbf{x}} \cdot \hat{\mathbf{x}}_g]\frac{T_k}{T}\right\}}{\sum_g [1+D(g)]\exp\left\{[-\frac{x^2}{2}D(g) + x\hat{\mathbf{x}} \cdot \hat{\mathbf{x}}_g]\frac{T_k}{T}\right\}}.$$

Further project the tensor equation of Eq. 5 into the $\hat{\mathbf{x}}$ direction, we can calculate the Curie–Weiss temperature by

$$\text{Eq. 9. } T_0 = T_k \frac{\sum_g \left[\frac{\hat{\mathbf{x}} \cdot \mathbf{\Phi}(g) \cdot \hat{\mathbf{x}}_g}{\hat{\mathbf{x}} \cdot \mathbf{\Phi}(0) \cdot \hat{\mathbf{x}}}\right]^2}{\sum_g [1+D(g)]}$$

If one takes $D(g) \equiv 0$ for the tetragonal ferroelectric material, one would get the Curie–Weiss temperature as $T_0 = T_k/3$ by Eq. 9.

We might point out that the above procedures can mostly recover the results of the mean-field approximation for the magnetic system if we treat $D(g) \equiv 0$. For instance, for spin ½ system, we simply get $T_0 = T_k$ and $x_{eq} = \tanh(x_{eq} T_k/T)$, which is the result of the Weiss molecular field theory [6]. Furthermore for spin ½ system, we want to add that at $T = 0$ K, our THF theory gives rise to $\Delta G = 0$ which is more reasonable than the nonzero value at 0 K given in most of the textbook [7]. This is especially important consider the fact that in most of the first principles calculation, the magnetic interaction is already accounted for in the total energy. If

$\Delta G$ were not treated as zero, then the nonzero value would be double-counted. The situation is the same for the excess enthalpy discussed in the next section, i.e., $\Delta H = 0$ at 0 K.

However, being beyond both the Weiss molecular field theory and the Bragg-Williams approximation, we want to point out that *both the type of lattice and the spatial dimensionality play the important roles* in the present THF description of the excess free energy as clearly seen from Eq. 6, knowing that fact that $\hat{\mathbf{x}}_g$ depends on both the type of lattice and the spatial dimensionality. This is particular true as we will see for the case of ferroelectric-paraelectric phase transition in PbTiO$_3$ in the following sections.

## IV. The Example Tetragonal System of PbTiO$_3$

Next let us handle the specific type of the tetragonal ferroelectric material PbTiO$_3$. We note that $\hat{\mathbf{x}}_g$'s for PbTiO$_3$ are three-dimensional and can take the values of (-1,0,0), (1,0,0), (0,-1,0), (0,1,0), (0,0,-1), and (0,0,1). By treating

Eq. 10, $\boldsymbol{\Phi}(0) = \begin{Bmatrix} \phi_{11} & 0 & 0 \\ 0 & \phi_{11} & 0 \\ 0 & 0 & \phi_{33} \end{Bmatrix}$

and $\hat{\mathbf{x}} = (0,0,1)$, Eq. 3 or Eq. 6 is reduced to

Eq. 11, $\Delta G(x,T) = \dfrac{k_B T_k}{2}(1+x^2) - k_B T \ln\left\{4\exp[-\dfrac{x^2}{2}D\dfrac{T_k}{T}] + 2\cosh(x\dfrac{T_k}{T})\right\}$.

where we have utilized that $D(g)$'s equal to a constant $D$ for the orientations (-1,0,0), (1,0,0), (0,-1,0), and (0,1,0) while $D(g)$'s equal zero along (0,0,-1), and (0,0,1). Similarly, for tetragonal system, Eq. 8 is reduced to

$$\text{Eq. 12, } x_{eq} = \frac{\sinh(x_{eq}\frac{T_k}{T})}{2(1+D)\exp[-\frac{x_{eq}^2}{2}D\frac{T_k}{T}]+\cosh(x_{eq}\frac{T_k}{T})}.$$

and the excess enthalpy (or internal energy) becomes

$$\text{Eq. 13, } \Delta H(x_{eq},T) = \frac{k_B T_k}{2}(1-x_{eq}^2) - k_B T_k x_{eq}^2 \frac{D\exp[-\frac{x_{eq}^2}{2}D\frac{T_k}{T}]}{2\exp[-\frac{x_{eq}^2}{2}D\frac{T_k}{T}]+\cosh(x_{eq}\frac{T_k}{T})}.$$

The calculated excess free energies by our THF model are plotted in Figure 1b to compare with those plotted in Figure 1a from the Landau theory. We observe that, our excess free energy decreases monotonically with increasing temperature, implying positive excess entropy, being opposite to those from the Landau model. Among the examined temperature range 314 K – 964 K, our model demonstrates that the highest excess free energy is that at 314 K of the ferroelectric phase. With a temperature step of 50 K, the excess free energy shows a monotonic decrease across the phase transition temperature at 764 K, finally down to the paraelectric phase with lowering free energy.

The calculated heat capacity, excess entropy, and excess enthalpy by our THF model are compared with those from the Landau model in Figure 2a, Figure 2b, and Figure 2c, respectively. For heat capacity, we see substantial improvement of the present THF mode over the Landau model, considering the experimental uncertainties seen by the difference between the measured heat capacity by Yoshida et al. [4] and Rossetti et al. [5]. It is observed that the critical behaviors of the excess entropy and excess enthalpy starting from 600 K approaching to the critical temperature are reproduced very well by the present THF model over the Landau model. As about the deviations of the THF model from the estimated points by Rossetti et al. [5] below 600 K, we might mention the possible uncertainties of procedure in extracting the excess thermodynamic quantities by Rossetti et al. [5]. Firstly, to obtain the excess heat capacity, Rossetti et al. [5] estimated the background or 'hard-mode' heat capacity by extrapolating the their measured heat capacity data in high temperature phase at T≥800 K to the data at T ≥ 450 K. Secondly, it is quite noticeable from Figure 2a the deviations between the heat capacity data measured by Yoshida et al. [4] and Rossetti et al. [5]

Last, let us examine the isothermal excess free energy of $PbTiO_3$ about the temperature evolution behaviors predicted by the THF model at proximity of the critical temperature in Figure 3. We see three characteristic temperatures, i.e., ~755 K, ~762 K, and 764 K. The value of 755 K represents the Curie–Weiss temperature $T_0$, calculated using Eq. 5 or Eq. 9 with the condition of the zero second order derivative of $\partial^2 \Delta G(\mathbf{X},T)/\partial \mathbf{X} \partial \mathbf{X} = 0$ at $\mathbf{X} = 0$. In the situation of $T = 762$ K, three local minima ($x = 0$ and $\pm 0.468$) are found along the excess free energy curve, i.e., there is a local, rather than global minimum at the point $x = 0$. This means that the phase transition predicted by the THF is first order as indicated by the fact that for the phase transition to occur, a discontinuous jump in the order parameter of $x$ is expected. The value of 764 K represents the critical temperature, at which the two local minima at $x \pm 0.468$ have been evolved into two shoulders at 764 K at the two finite values of $x = \pm 0.385$, from which a jump of $x$ into zero to be happened to have the phase transition to complete.

V.  **Remarks**

In summary, we have presented a thermal harmonic field description of a system under phase transition based on *a priori* assumption that the macroscopic system is made of the thermal mixing among multi harmonics each of them can be distinguished by crystal orientation, polar direction, magnetic direction, or even momentum etc. Several important improvements over the widely adopted Landau theory include: i) the excess free energy is well formulated using the second order term of the order parameter, compared with those use up to the sixth order term by the Landau theory; ii) the excess free energy is referenced to the low temperature structure whose free-energy can be well formulated and calculated within the modern first-principles theory, compared with that referenced by the Landau theory to the high temperature structure which shows imaginary phonon modes; iii) the excess entropy is always positive resulting in that the change of free-energy always decreases with increasing temperature, compared with negative excess entropy by the Landau theory; iv) the long range field is naturally derived; v) for electric polar system, ferroelectric-paraelectric phase transition and associated thermodynamic behaviors are more accurately described.

Last, we want to remark the effect of the external field can be considered by the following modification of Eq. 2

Eq. 14
$$\Delta F(\mathbf{X}, \mathbf{E}; g) = \frac{1}{2}[\mathbf{X} - \mathbf{X}_0(g)] \cdot \mathbf{\Phi}(g) \cdot [\mathbf{X} - \mathbf{X}_0(g)]$$
$$+ \frac{1}{2}\mathbf{E} \cdot \mathbf{\Psi}(g) \cdot [\mathbf{X} - \mathbf{X}_0(g)] + \frac{1}{2}[\mathbf{X} - \mathbf{X}_0(g)] \cdot \mathbf{\Psi}(g) \cdot \mathbf{E}$$
$$+ \frac{1}{2}\mathbf{E} \cdot \mathbf{K}(g) \cdot \mathbf{E}$$

However, we want to leave the work to the future.

VI.   **Calculational details**

For completion, first-principles phonon calculations under the quasiharmonic approximation, employing the mixed space supercell phonon approach [8], have been performed to consider the phonon effect of the lattice vibration in PbTiO$_3$. The value of $T_k$ in Eq. 6 is taken as $T_k = 2388$ K. As about the evaluation of the parameter $D$ in Eq. 11, we assume that $\Phi(0)$ in Eq. 10 is proportional to the high frequency static dielectric tensor $\boldsymbol{\varepsilon}^\infty$. With the calculated vaule of $\varepsilon_{11}^\infty = 7.594$ and $\varepsilon_{33}^\infty = 7.029$ calculated using the first-principles Berry-phase approach of Nunes and Gonze [9], we get $D = \varepsilon_{11}^\infty / \varepsilon_{33}^\infty - 1 = 0.0804$.

One further factor needs to be considered in the present THF model is the size of the SSCD. By the rough value of the excess entropy of ~5.68 J/mole K$^{-1}$ for the ferroelectric-paraelectric phase transition as estimated by Rossetti et al. [5] based on their experimental heat capacity data, compared with an ideal disordering among the six orientation directions in the form of $R\log 6 = 8.31446 \times 1.79176 = 14.8975$ 5.68 J/mole K$^{-1}$ ($R$ is the gas constant), we simply assume each SSCD contains 2 formula unit of PbTiO$_3$, i.e, $N = 2$ has been used in all the THF results plotted from Figure 1 to Figure 3. We do not try to best fine-tune the value of $N$ considering the quite noticeable deviations between the heat capacity data measured by Yoshida et al. [4] and Rossetti et al. [5] as shown in Figure 2a.

For first-principles calculation, we have employed the projector-augmented wave (PAW) method [10] implemented in the Vienna *ab initio* simulation package (VASP, version 5.2). The exchange-correlation functional according to Ceperley and Alder as parameterized by Perdew and Zunger [11] was employed. The Born effective charge tensor and the high frequency static dielectric tensor needed in the phonon calculation are calculated employing the linear-response theory implemented in VASP 5.2 by Gajdos et al. [12] within Berry-phase approach of Nunes and Gonze [9].

ACKNOWLEDGEMENTS

This paper was written with supports of the U.S. Department of Energy under Contract No. DE-FC26-98FT40343 (Wang and Liu) and the DOE Basic Sciences under Grant No. DOE DE-FG02-07ER46417 (Wang and Chen). The simulations were carried out on computer clusters at Penn State partially supported by instrumentation funded by the National Science Foundation through grant OCI-0821527. Some of the calculations were performed at the National Energy Research Scientific Computing Center, which is supported by the Office of Science of the U.S. Department of Energy under Contract No. DE-AC02-05CH11231.

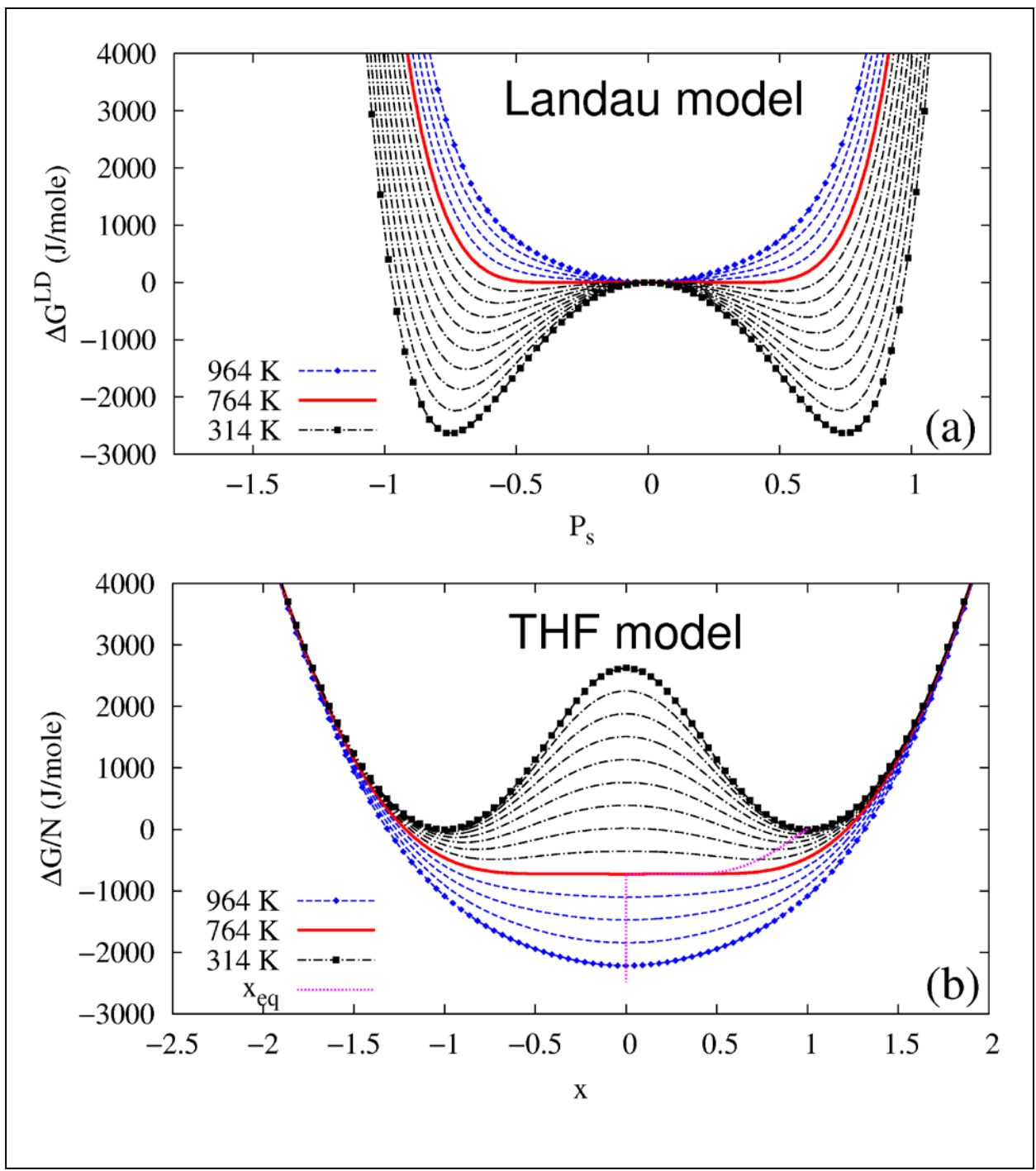

Figure 1. Temperature evolutions in the temperature range of 314 K – 964 K of the isothermal excess free energy of PbTiO$_3$ as a function of the order parameter. The temperature step between the neighboring curves is 50 K. (a) results from the Landau model using the spontaneous

polarization as the order parameter and (b) results the present THF approach where the the dashed (pink) line marks the global minima along the isothermal curves. The solid (red) line is for the critical temperature of 764 K; the nine dot-dashed (black) lines are for temperatures lower than 764 K; and the four dashed (blue) lines are for temperatures higher than 764 K.

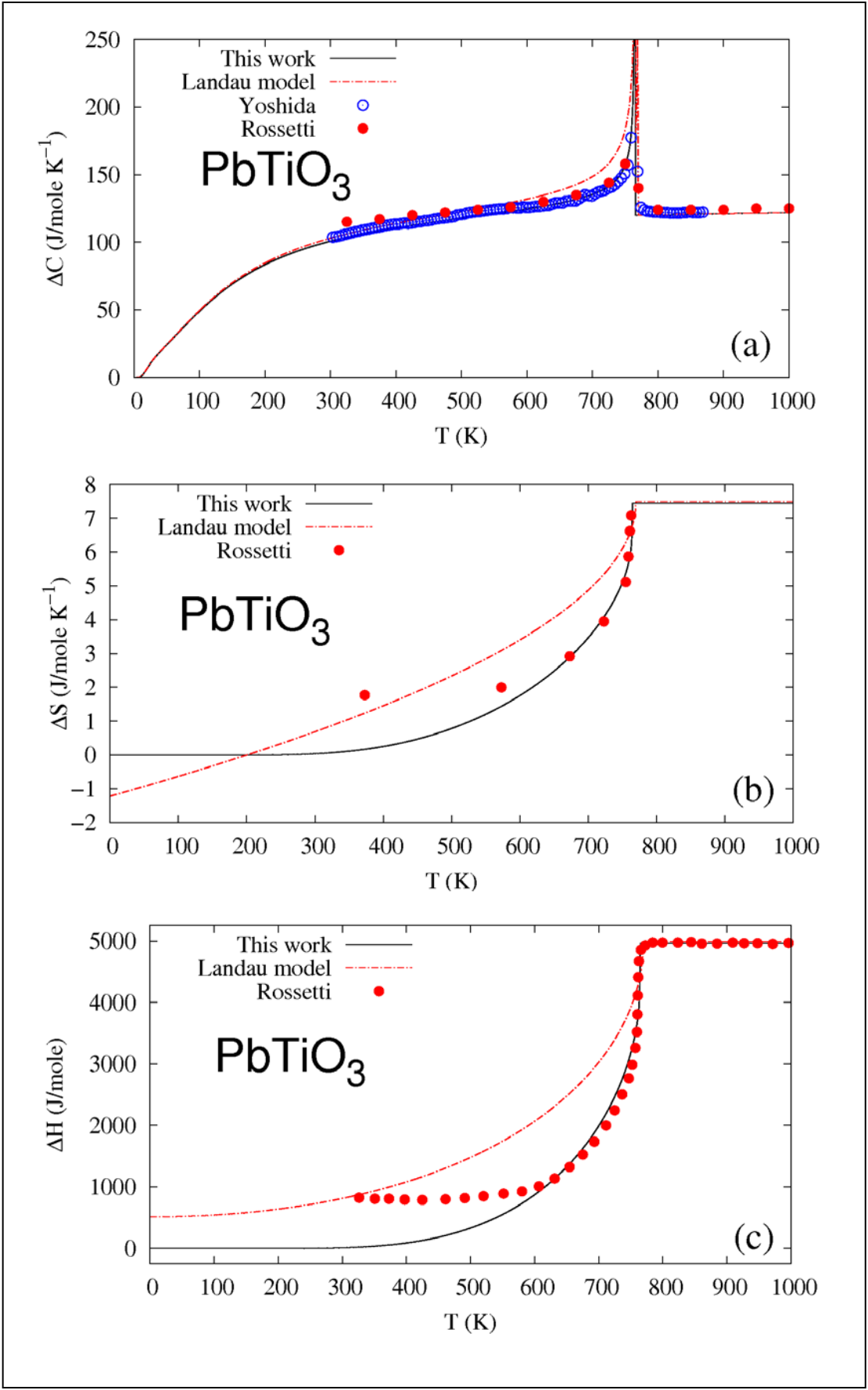

Figure 2. The evaluations of the Landau model (dot-dashed lines) and the present harmonic fluctuation field approach (solid lines) in terms of the thermodynamic properties of $PbTiO_3$. (a) heat capacity, where the open circles are experimental data measured by Yoshida et al. [4] and the solid circles are experimental data measured by Rossetti et al. [5].; (b) excess entropy, where the solid circles are the estimated data by Rossetti et al. [5] based on their experimental heat capacity.; and (c) the excess enthalpy, where the solid circles are the estimated data by Rossetti et al. [5] based on their experimental heat capacity. In (b) and (c), the results from the Landau model and the estimated data by Rossetti et al. [5] have been uniformly shifted to align with the present THF results at the paraelectric phase for the purpose of comparison.

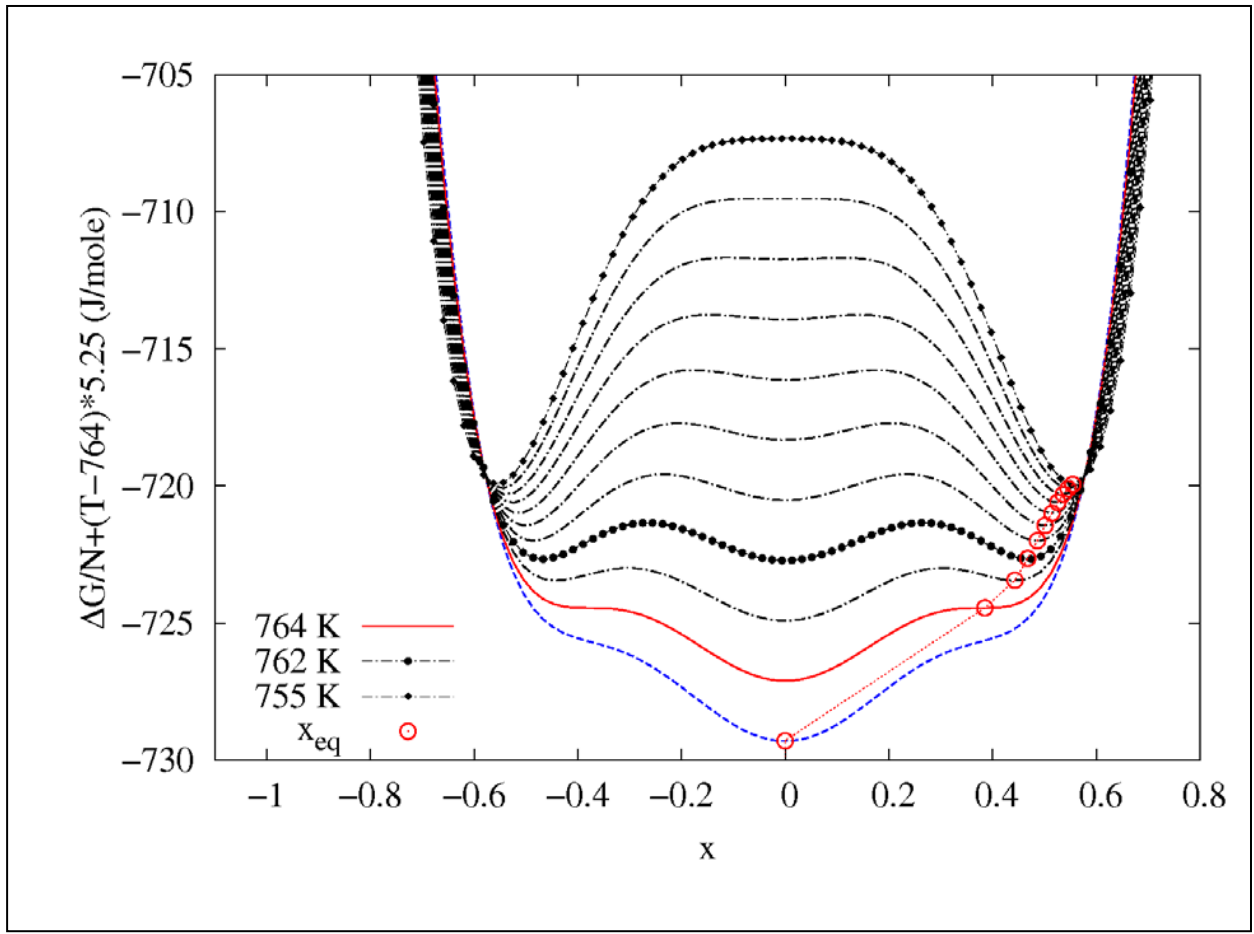

Figure 3. Temperature evolutions of the isothermal excess free energies of PbTiO$_3$ in the range of 755 K – 765 K. The temperature step between the neighboring curves is 1 K. A shift of in the amount of $5.25 \times (T - 764)$ J/mole have been applied to the plotted curves. The dot-dashed line with the solid diamond points represents the isothermal at the Curie–Weiss temperature $T_0 = 755$ K; The dot-dashed line with the solid circle points shows the isothermal for $T = 762$ K at which three local minima are seen (x = 0 and ±0.468) ; the solid (red) line shows the isothermal for the critical temperature $T = 764$ K at which two shoulders are seen at x = ±0.385; and the dashed (blue) line shows the isothermal for $T = 765$ K at which only one global minimum at x = 0 is seen.